\documentstyle[psfig]{l-aa}

\begin{document}

\thesaurus{  
       (
        12.12.1;  % (Cosmology:) large-scale structure of Universe
        11.03.1);  % Galaxies: clustering
          }

\title{Count laws and projection effects in clusters of galaxies 
       \thanks{Based on observations collected at the ESO (La Silla, Chile) 
and CFHT}
      }

\author{C.~Adami \inst{1}, P.~Amram \inst{2},  G.~Comte \inst{2}}
\institute{IGRAP, Laboratoire d'Astronomie Spatiale, Marseille, France 
   \and IGRAP, Observatoire de Marseille, France }

\offprints{ C.~Adami}
\date{Received date; accepted date}

\maketitle
\markboth{ Count laws and projection effects in clusters of galaxies }{}

\begin{abstract}
We show that a 2D projection is representative of its corresponding 3D 
distribution at a confidence level of 90 $\%$ if it follows a King profile 
and if we consider the whole spatial distribution. The level is significantly 
lower and not decisive in the vicinity of the 2D cluster center. On another 
hand, if we verify the reciprocal statement of the Mattig's distribution 
(1958) -i.e. a flux limited sample is represented by a 0.6 slope of its 
count law-, we point out that, due to the usual unaccuracy of the slope
determination, a slope of 0.6 is not a sufficiently strict criterion for 
completeness and uniformity of a sample as often used in the literature.
\end{abstract}

\begin{keywords}
{ (Cosmology:) large-scale structure of Universe
-Galaxies: clustering
}
\end{keywords}

\section{Introduction}

A large part of modern observational cosmology focuses on statistical
studies of the galaxy population in the Universe. Two problems are 
commonly faced that concern the properties of galaxy samples:

First, the spatial distribution of galaxies belonging to clusters is basically 
unknown. The image of a cluster is a 2-dimensional projection of the true 
3-dimensional distribution, and one is led to question the quality of this 2D 
projected distribution as estimator of the true 3D one. Let us recall that this 
point is important for understanding the true physics of the cluster, and that 
properties as the core radius (Lubin \& Postman 1996) and morphological 
segregation (e.g. Whitmore \& Gilmore 1991, Stein 1996) are derived from 
the 2D distributions only.

Second, one usually uses the slope of the count law of flux-limited 
extragalactic samples to estimate their completeness (e.g. Paturel et al. 
1994) and uniformity. This test 
is based on Mattig's demonstration (1958) that, in an Euclidian Universe, the 
number $N$ of objects brighter than an $m$ magnitude is: $N(m)
\propto 10^{0.6m}$. However, the validity of the reciprocal statement, that 
a count law with slope 0.6 in ($m, log N$) coordinates implies uniformity 
and completeness, remains to be checked carefully.

The present paper aims at clarifying these two questions by means of 
Monte-Carlo simulations based on input parameters consistent with observed ones.
For this purpose, a serie of test sample, containing either simulated and real
data have been considered. For this purpose, series of test samples containing 
either simulated or real data, have been considered.

The samples are presented in Section II. Section III describes the relationship
between the projected 2D and real 3D distributions of galaxies in clusters. 
Section IV discusses the slope of counts laws and completeness. Section V 
contains our conclusion and summary.
 
\section{Description of the samples}
\subsection{Density laws}

\begin{itemize}
\item  
To check projection effects in clusters of galaxies, two projected density laws 
$\rho (r)$ have been considered:

an uniform one with a constant density

a generalized 2D King law of $\beta$ exponent with core radius $r_c$:
$\rho (r)\propto [1/(1+ (\frac{r}{r_c}) ^2)] ^\beta $

From these distributions, 3D spherical distributions are generated by 
affecting to each particle (galaxy) a radial coordinate $z$ using a 
gaussian random generator. We recall that the core radius $r_c$ is 
the same for a 2D or a 3D King profile (Girardi et al. 1995).
\end{itemize}

\begin{itemize}
\item  
To check the validity of the reverse count law, we generate 3 artificial
spherical random distributions. These populations of galaxies are based
on uniform space density with various clustering structure. 
\end{itemize}

\subsection{Cluster samples}
We used also four samples taken from 52 clusters of galaxies:

The first one (Sample I) is built from Amram et al. (1992, 1994, 1995) 
samples containing 34 spiral galaxies taken from 8 clusters. To be more 
complete, other morphological type galaxies have been added in order 
to follow a King profile according to Whitmore \& Gilmore (1991). The 
core radius of this synthetic cluster is 150 kpc.

The second (Sample II) is a sample of 43 Edinburgh-Durham Southern Galaxy
Catalogue (hereafter EDSGC) and Eso Nearby Abell Cluster Survey (hereafter 
ENACS) rich Abell clusters (table 1) kindly provided by ENACS and COSMOS 
team. These clusters are 
described at length in Mazure et al. (1996), in Katgert et al. (1997), and in 
Heydon-Dumbleton et al. (1989). EDSGC catalog gives the projected 
distribution of all galaxies. The EDSGC data are nearly complete for 
$b_j$=20. Core radii, exponent $\beta $ and center for these 43 clusters are 
known (see Adami et al. 1997: AMK98 hereafter). We have selected here 
only nearly circular clusters to minimize ellipticity effect (table 1). We 
take into account only galaxies in an homogeneous zone extending across  5 
core radii. This corresponds to the main part of each cluster.

The third sample (Sample III) is taken from Virgo cluster. We have more
than 550 galaxies in the whole available area (Binggeli et al. 1985). We do
not have the density law information but this cluster will be used only for
uniformity and completeness tests.

\subsection{Artificial samples to test reciprocal Mattig's law}

To check the validity of the reverse count law assumption, we generate in
an Euclidean universe spatial distributions of objects with various uniformity 
and apparent magnitude completeness conditions. We assume that a sample is 
complete if it follows in absolute magnitude a Schechter luminosity function 
with de Lapparent parameters (1989). The range of generated absolute magnitude 
is [-11,-23].

The first one (Sample IV) has no structure in its space distribution, the 
position of each galaxy inside the volume is purely random. The uniform 
density is fixed  to 0.06 galaxies Mpc$^{-3}$ consistently with 
a projected surface density of 3 $10^{-5}$ $gal~arcsec^{-2}$ (Colless 
1989). For computing limitation reasons, we populate only the 1/8 fraction 
of a spherical volume of  300 Mpc radius (H$_0$=100 km.s$^{-1}$Mpc
$^{-1}$ and q$_0$=0) with 6.55 10$^6$ galaxies. The 1/8 fraction does not 
affect the validity of the results. The radius of the sphere corresponds to a 
redshift lower than 0.1  i.e. to a space volume 
within which the Universe is assumed to be quasi-euclidian. The galaxy 
absolute magnitude distribution is complete according to a Schechter 
luminosity function with $\alpha $=1.1 and $M_{*}$=-19.2 (de Lapparent 
et al. 1989). 

In the second sample (Sample V), the galaxies are grouped into 50 clusters of 
identical population. Each of these clusters is a sphere of 1 Mpc diameter. 
Two spheres are in average separated by 66 Mpc. The position of the cluster 
centers and the position of the galaxies inside the clusters are randomly 
generated. This mimics a distribution of galaxies in discrete groups bathing 
in large voids.

Sample VI combines a cluster distribution analog to Sample V case, 
encompassing 2/3 of the total galaxy population with a remaining 1/3
randomly distributed as isolated objects across the whole volume. The ratio 
2/3 and 1/3 are given by Combes et al. (1991).

\begin{table}
\caption[]{The 43 clusters of galaxies with the correlation coefficient
$C$, the representiveness confidence level of the mean value of $C$, 
and the slope of the count law before truncation (Slb) and after 
truncation (Sla).}
\begin{flushleft}
\small
\begin{tabular}{rcrrr}
\hline
\noalign{\smallskip}
name & $C$ & $P$ & Slb & Sla \\ 
\noalign{\smallskip}
\hline
\noalign{\smallskip}
A0087 & 0.68$\pm $0.08 & 95 & 0.48  & 0.48 \\ 
A0118 & 0.70$\pm $0.08 & $\geq$ 99 & 0.62  & 0.43 \\ 
A0119 & 0.68$\pm $0.07 & $\geq$ 99 & 0.45  & 0.39 \\ 
A0151 & 0.67$\pm $0.09 & $\geq$ 99 & 0.54  & 0.45 \\ 
A0168 & 0.67$\pm $0.05 & $\geq$ 99 & 0.47  & 0.43 \\ 
A0367 & 0.62$\pm $0.07 & 72 & 0.58  & 0.55 \\ 
A0524 & 0.67$\pm $0.05 & 82 & 0.58  & 0.50 \\ 
A0978 & 0.72$\pm $0.11 & $\geq$ 99 & 0.64  & 0.56 \\ 
A1069 & 0.67$\pm $0.04 & 97 & 0.44  & 0.40 \\ 
A2353 & 0.65$\pm $0.08 & 86 & 0.69  & 0.64 \\ 
A2362 & 0.63$\pm $0.07 & 77 & 0.31  & 0.29 \\ 
A2383 & 0.71$\pm $0.11 & $\geq$ 99 & 0.57  & 0.47 \\ 
A2426 & 0.70$\pm $0.08 & $\geq$ 99 & 0.63  & 0.53 \\ 
A2480 & 0.68$\pm $0.07 & 81 & 0.51  & 0.46 \\ 
A2644 & 0.68$\pm $0.09 & 83 & 0.43  & 0.30 \\ 
A2715 & 0.71$\pm $0.09 & $\geq$ 99 & 0.61  & 0.57 \\ 
A2717 & 0.71$\pm $0.03 & $\geq$ 99 & 0.52  & 0.47 \\ 
A2721 & 0.68$\pm $0.04 & $\geq$ 99 & 0.79  & 0.67 \\ 
A2734 & 0.66$\pm $0.06 & $\geq$ 99 & 0.46  & 0.35 \\ 
A2755 & 0.59$\pm $0.10 & 54 & 0.64  & 0.54 \\ 
A2764 & 0.65$\pm $0.07 & 93 & 0.55  & 0.51 \\ 
A2765 & 0.69$\pm $0.13 & 96 & 0.62  & 0.50 \\ 
A2778 & 0.67$\pm $0.08 & 98 & 0.62  & 0.52 \\ 
A2871 & 0.60$\pm $0.07 & 61 & 0.43  & 0.28 \\ 
A2911 & 0.70$\pm $0.06 & $\geq$ 99 & 0.29  & 0.29 \\ 
A2923 & 0.59$\pm $0.09 & 52 & 0.45  & 0.34 \\ 
A3093 & 0.59$\pm $0.12 & 53 & 0.65  & 0.53 \\ 
A3094 & 0.68$\pm $0.05 & $\geq$ 99 & 0.66  & 0.55 \\ 
A3111 & 0.65$\pm $0.06 & 90 & 0.33  & 0.30 \\ 
A3122 & 0.68$\pm $0.04 & $\geq$ 99 & 0.66  & 0.53 \\ 
A3128 & 0.67$\pm $0.02 & 90 & 0.60  & 0.51 \\ 
A3151 & 0.66$\pm $0.09 & 88 & 0.54  & 0.47 \\ 
A3158 & 0.68$\pm $0.07 & $\geq$ 99 & 0.57  & 0.49 \\ 
A3194 & 0.71$\pm $0.06 & $\geq$ 99 & 0.64  & 0.55 \\
A3341 & 0.66$\pm $0.09 & $\geq$ 99 & 0.51  & 0.43 \\ 
A3528 & 0.62$\pm $0.04 & 82 & 0.40  & 0.35 \\ 
A3744 & 0.69$\pm $0.06 & $\geq$ 99 & 0.54  & 0.49 \\ 
A3781 & 0.65$\pm $0.09 & 84 & 0.45  & 0.43 \\ 
A3809 & 0.68$\pm $0.03 & $\geq$ 99 & 0.68  & 0.58 \\ 
A3822 & 0.69$\pm $0.03 & $\geq$ 99 & 0.77  & 0.63 \\ 
A3825 & 0.68$\pm $0.06 & $\geq$ 99 & 0.60  & 0.54 \\ 
A3827 & 0.70$\pm $0.06 & $\geq$ 99 & 0.69  & 0.55 \\ 
A3897 & 0.66$\pm $0.10 & 90 & 0.59  & 0.55 \\ 
\hline	   
\normalsize
\end{tabular}
\end{flushleft}
\label{t-data1}
\end{table}

\section{2D/3D representativeness}

To check the consistency between projected 2D and spatial 3D distributions of 
galaxies in a cluster, we compute the Pearson correlation coefficient $C$ 
between the projected radius $r$ and the real radius $s=\sqrt{r^2+z^2}$ 
for the whole sample. We have $C$=$\frac {cov(r,s)}{\sqrt{var(x)var(y)}}$.

In order to quantify the signification of $C$, we generate artificial 2D 
distributions and their corresponding 3D ones with exactly the same number of
objects than the observed clusters, but without any 2D structures. We impose 
a low ellipticity of about 0.1 along the line of sight. This is the mean 
observed value of the projected ellipticities of our real clusters (AMK98). 
We have then for the artificial cases the lowest possible level of 
representation 
between a 2D and a 3D structure. For a given observed cluster, we proceed 
10000 realizations of the artificial structure. By this way, we are able to 
calculate the percentage $P$ of realizations (without structure) which give 
a lower value of $C$ than the observed case (with structure). P is called 
here the confidence level of $C$. 

For Sample I, we generate 10000 3D distributions according to a King density 
law with r$_c$ = 150 kpc and $\beta $=1. We find C=0.70$\pm$0.15. The 2D
distribution is representative of the 3D one with a confidence level of more 
than 99$\%$. However, this sample contains less than 200 galaxies and is 
partially ideal. To deal with more realistic clusters, we use Sample II.

For Sample II, we generate 1000 3D distributions for each 
cluster according to a King density law with the parameters derived for each 
cluster by AMK98. The average $C$ value is 0.67 for Sample II. 
The standard error on each real cluster is typically 0.07 and the standard 
error of mean $C$ determination on all clusters is 0.03. We note that the 
synthetic Sample I and the real Sample II give very consistent values of  $C$.
We give in table 1 the level $P$ for each cluster. As we can see, only 10 
clusters show a percentage lower than 90$\%$. The 33 others are significantly 
representative of the 3D distribution at a level of more than 90$\%$.

\begin{figure}
\vbox
{\psfig{file=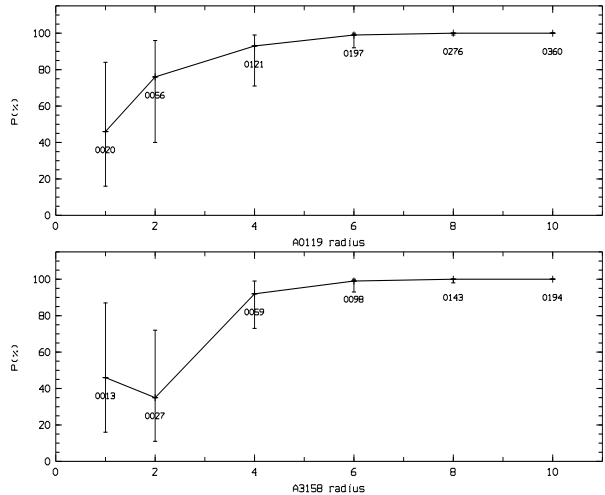,width=8.7cm,height=7.0cm,angle=270}}
\caption[]{Variation with the normalized radius $D_i$ of the representativeness 
confidence level for A0119 and A3158. The number of galaxies within each radius 
is labeled.}
\label{comp}
\end{figure}

To check this representativeness variation as a function of the included area, 
we use 2 specific clusters, member of Sample II, that are very regular and 
contain more than 200 galaxies: A0119 and A3158. This ensures good statistics. 
Applying  the same 3D generation method, we consider 5 values of the ratio 
$D_i$=$\frac{i^{th} {limiting~cluster~radius}}{core~radius}$. We 
impose $D_i$ = 1, 2, 4, 6 and 8. We calculate the correlation
coefficient $C$ and its representative confidence level P for each 
case (Fig.~1). $P$ increases from about 50 $\%$ with large error bars to more
than 99$\%$. The central distributions are significantly less constrained than 
more extended ones and are only poorly representative of the corresponding 3D 
ones. We note here, that this tendency is certainly dependent on the 
actual cluster spherical degree: a strongly elongated cluster with a major axis 
along the line of sight will exhibit a lower P. Moreover, this 
representation is directly related to the morphological type of the galaxies:
we know that the spirals are significantly less concentrated than the 
ellipticals (e.g. Adami et al. 1998 and included references: ABM98).

\section{Slope of count laws and completeness}

\subsection{Non uniform samples}

For each of the Samples IV, V and VI, 100 Monte-Carlo runs are 
done. The observed counting laws, after projection into apparent magnitudes, 
are the following: 

\begin{itemize}
\item  
Sample IV, as expected, gives a slope 0.59 $\pm $ 0.02, consistent with 
Mattig's (1958) law.
\end{itemize}

\begin{itemize}
\item  
Sample V gives a slope 0.89 $\pm$0.18, showing that extreme non-uniformity 
translates into a much steeper relation than Mattig's one and a 
non-reproducibility of the slope value, as evidenced by the statistical 
dispersion.
\end{itemize}

\begin{itemize}
\item  
Sample VI, supposed to be a more realistic picture of the nearby Universe 
than  the 
purely uniform case, yields a slope 0.64$\pm$0.04, only slightly departing 
from Mattig's 
value. Therefore, a substantial degree of inhomogeneity in the distribution of 
galaxies 
should be quite difficult to detect from observation of the counting law slope 
only.
\end{itemize}

We use our available Abell clusters (Sample II) as the "real" sample. We
have calculated the slope of the count law in the range [minimal cluster
magnitude, minimal cluster magnitude + 3] for the largest available area of
each cluster. We find a slope equal to 0.56$\pm $0.11. We note that the
error on each slope is negligible compared to the global dispersion of the 
individual values. This mean value is consistent with the uniform case: 14 
of the clusters show a slope greater than 0.56.
We remark that one of the most non uniform clusters, like A0151, provides 
a slope of 0.54 close to the ''uniform reference case'' value of 0.59.

We do not have $very$ different values of the slope between purely
uniform samples and samples with a uniform part or real samples. The only
significant discrepancy occurs when considering a totally non uniform sample, 
which is not observed, as soon as a survey encompasses a sufficient solid 
angle.

\subsection{Effect of incompleteness}

First, the slope of the count law of the Virgo sample (Sample III) is analyzed
in various apparent magnitude ranges. It is found to be equal to 0.59$\pm$0.09 
in [9.0;12.2]. Hence, completeness is taken as working hypothesis across this 
range. Further, we cut
randomly with a uniform law of the magnitude about 50 \% of the objects and 
we recalculate the slope of the count law. We find 0.58$\pm$0.10, not different 
from the complete case.

We use now the structureless Sample IV and we cut it in a more realistic way 
according to the exponential law of the magnitude m: 1-e$^{(m-(min+2))/
(m-max)}$. The variable $min$ is the minimum observed magnitude in the sample 
and $max$ is the theoretical limiting magnitude of the EDSGC survey. We deal 
now with appromatively 1.07 10$^6$ galaxies out of the original 6.55 10$^6$ 
ones. We find a slope of 0.53$\pm$0.02 to be compared with 0.59 $\pm$0.02 
(see previous section). The difference is significant, but we must take into 
account the very small errors due to the very large number of considered 
galaxies. 

Third, we use the Abell clusters sample (Sample II). We cut their galaxy 
apparent magnitude distribution with the same exponential law .When $m$ grows, 
the percentage of removed objects grows also. We reject by this way about 
70$\%$ of the galaxies. We compare the calculated slopes before and after 
truncation (table 1). The mean truncated value is 0.48$\pm$0.09 and the mean 
difference between the two slopes is 0.07 (15\% of the slope), lower than 1 
standard error. More than 30\% of the clusters exhibit a difference smaller 
than 10\%. This difference represents an error lower than the usual accuracy 
of the slope determination (when the counting errors are estimated with the 
usual square root of the count). We note however that the truncated values are 
systematically lower than the non truncated ones.

\section{Conclusion}

We have shown that a 2D projection is representative at a level of 90$\%$ 
of the corresponding 3D sample if it obeys a King profile and if we consider 
the whole cluster. The confidence level is significantly lower in the vicinity 
of the cluster 2D center (typically inside 4 core radii) and is not decisive 
there.
 
We have also shown that it is somewhat unsafe to use the slope of the count law 
of a sample to test its completeness level and its uniformity. Indeed, for the
uniformity of the sample, we have almost no observable difference between a 
realistic artificial sample (with 2/3 of its population in clusters), and a 
completely uniform artificial sample. 
For the completeness level, the difference is about 15\% between a complete 
sample and an incomplete one. This difference represents an error lower than 
the usual accuracy of the slope determination. We conclude that the slope of 
the count law, in itself, is not a decisive factor to assess uncompleteness and 
homogeneity of a sample.

\begin{acknowledgements}

{The authors thank A. Mazure, all the ENACS team and EDSGC members for the 
use of EDSGC data and for helpful discussions.}

\end{acknowledgements}

\vfill
\end{document}